\documentstyle[12pt]{article}
\input psfig.tex
\begin{document}
\begin{center}
{\Large\bf Correlation of Ultra High Energy Cosmic
 Rays with Compact Radio Loud Quasars}

\vskip 1cm

{\large Amitabh Virmani$^a$, Sibesh Bhattacharya$^a$,
Pankaj Jain$^a$,\\ 
Soebur Razzaque$^b$, John P. Ralston$^b$ and 
Douglas W. McKay$^b$}

\vskip 1cm
$^a$ Physics Department \\ I.I.T. Kanpur, India 208016\\
$^b$ Department of Physics \& Astronomy\\ University of
Kansas\\ Lawrence, KS 66045, USA
\end{center}

\vskip 1 cm
%PACS:   96.40, 98.70.S

%Keywords: Cosmic Rays, Cosmic Ray Sources, QSO, 
%Statistics

\begin{abstract} Angular correlations of ultra-high energy cosmic rays
with cosmologically distant sources may provide clues to these
mysterious events.  We compare cosmic ray tracks with
energies above $ 10^{ 20}eV$ to a compilation of radio-loud compact
QSO positions.  The statistical method emphasizes invariant quantities
and a test of {\it statistical independence} of track and source distributions.
Statistical independence is ruled out
by several independent statistics at confidence levels of less that
$10^{-3}$ ($99.9\% $.), indicating that track directions and QSO
source positions are correlated at a highly significant level.

\end{abstract}

\section{Introduction}
The origin and propagation of the highest energy cosmic rays 
\cite{linsley,agasa1999,fliesI,HP,winn} present a
major challenge to current understanding of astrophysics.  The problem
has developed over many decades.  In 1966 Greissen, Kuzmin and
Zatsepin ($GZK$)\cite{GZK} discovered a theoretical upper limit on cosmic ray
energies of about $4 \times 10^{ 19 }eV$.  The limit is due to
laboratory-established pair and photo-nuclear production processes
occuring on the cosmic background radiation.  Due to these processes,
protons of such energies are unable to propagate distance greater than
about 50 Mpc.  Other nuclei are even more severely restricted \cite{puget}.  
It is
commonly believed \cite {Hillas,blandford,nagano} 
that the only possible astrophysical
sources of particles with such high energies are the active galactic
nuclei ($AGN$).  The vast majority of $AGN$ are cosmologically
distant.  There are insufficient alternative sources within 50 Mpc of
Earth to produce the observed events.  Hence the origin of numerous
events above $4 \times 10^{19 }eV$ is very puzzling.

The angular distribution of so-called ``$GZK$-violating'' events may
contain important clues.  In searching for a possible source for the
Fly's Eye event FE320 ($320 EeV$), Elbert and Sommers \cite
{ElbertandSommers} noticed that its arrival direction was very close
to the remarkable quasar 3C147.  Farrar and Biermann 
($FB$)\cite{Farrar98} pointed out
that 3C147 is a compact quasar with jets about one-tenth the size of a
full-sized quasar with radio lobes. The spectrum of 3C147
is cut off at low radio frequencies, providing another characteristic
of its compactness.

Here we re-examine possible correlation of compact radio-loud quasars
and cosmic-ray track directions.  $FB$ claimed a correlation between
track directions and compact quasars, defined by the following three
criteria:
\begin{itemize}
\item[1] The quasar should be listed in the NASA/IPAC extragalactic database
(NED).
\item[2] The object should be radio loud. In practice $FB$ required that
the object appear in the  K\"{u}hr catalog \cite{kuehr}.
This catalog contains a total 1835 radio sources including all those whose
flux density at 5 GHz is $\ge$ 1 Jy.
\item[3] The object should have flat or falling spectrum at low radio
frequencies.
\end{itemize}

We organize our study somewhat differently from previous work.  First,
we pay extra attention to the baseline of the statistic, namely the
definition of ``no signal'', employing several independent methods.
This makes the study more reliable and more conservative than previous
ones.  Second, we employ invariant quantitites, both for the purposes
of conceptual correctness and also to correct substantial relative
systematic errors.  Third and most importantly, we make no attempt to
validate any proposed correlation.  In fact, correlations can exist in
such myriad forms that any statistical procedure accepting one form
over another {\it a posteriori} can be suspect.

Of pivotal importance is that any proposed relation of
$GZK$-violating cosmic rays and cosmological sources should test a
crisp hypothesis.  The default for cosmologically-distant sources is a
relation of {\it statistical independence} of the arrival directions
relative to the direction of the sources.  Besides the $GZK$
attenuation, scrambling of directions by intergalactic fields also
tends to wash out any correlation of source-track directions.
Independence is a well-defined hypothesis which can be tested without
introducing model-dependence or extraneous postulates for these
puzzling events.

\section{More on the Data, and Analysis}

There is a total of 285 quasars in the K\"{u}hr catalog which have a flat
or falling spectrum at low radio frequencies.  The  
catalog contains no quasars within $\pm 10$ degrees of the galactic
plane.  
Furthermore the density of quasars in the catalog is much
smaller in the southern hemisphere than the northern
hemisphere.

Candidates displaying spectra similar to quasars but not classified as
quasars were ignored.  If all sources with flat or falling spectrum at
low radio frequencies were included, then the total number of sources
would be about 500, close to the number cited by Farrar and Biermann.
We will find that the remaining sources, not classified as quasars, do
not show any correlation with track directions in our data set.

There are 25 track events available with energies exceeding $10^{
20}eV$, a value chosen to be well beyond the $GZK$-violating regime.
The events are listed in Table 1.  We exclude those events which have
galactic latitude $\le$ 10 degrees since the catalog also
imposes this cut.  The cut removes 7 of the 25 events, leaving 18
tracks to be compared to the catalog QSO sample.

Rather than attempting to explain correlation, our strategy is to test
independence.  Independence of tracks and sources is expressed by the
joint distribution $$f(tracks, sources)= f(tracks) f(sources), $$
where we follow the usual practice of labeling a distribution by its
argument.  The coordinates of tracks and sources has previously been
taken to be the right ascsension and declination.  Using coordinates
as statistics can
introduce a human bias, namely the coordinate system of $RA, DEC$.  To
make an invariant statistic, we map each track and source into unit
vectors $\hat x_ { track } , \hat x_ {source}$ on the surface of the
celestial sphere.  We then look at the distribution of $\hat x_ {
track } \cdot \hat x_ {source},$ which being invariant does not
depend on the coordinate system.  The angle $\gamma$ between each
track ($i$) and source ($j$)is defined by $$ \gamma_{ ij } = cos^ {
-1 }(\hat x_ { track } \cdot \hat x_ {source}).$$  One can also
interpret $\gamma$ as
the minimum geodesic distance along the unit sphere between the two objects,
removing all reference to the astronomical coordinates.

To incorporate the experimental errors, we examine $\gamma^2$ in units
of the reported errors.  The Akeno Giant Air Shower Array (AGASA)
group reports an error cone 
containing 68\% of the events.
An error cone is ideal because it is also an invariant concept.
In Ref. \cite{cluster} the authors use an angular error of
$1.8^o$ for the AGASA and $3^o$ for the rest of the detectors.
We also adopt these error values for our analysis and later examine
how our results change if we allow the errors to change slightly.
The only exception to this is the Fly's Eye event where the error in
DEC and RA are reported as given in Table 1.
Letting
the particular error for each event be denoted $\delta \gamma$, we
create a statistic $$\delta\chi^2_i = min_j (\gamma_ { ij }
^2/\delta\gamma_i^2) .$$  Our $\delta\chi^2_i$ is the analogue of $FB$'s
statistic $\delta\chi^2_ { FB } = min_j ( ( RA_i-RA_j)^2 +(
DEC_i-DEC_j)^2)/\delta\gamma_i^2, $ which is not invariant due to the
curvature of the sphere.  While one might not expect problems
when angles are small, the RMS {\it relative} deviation of the two measures
  (in equatorial coordinates) is about 1.46 in the data set.

The distribution of $\delta\chi^2$ in the regions of very large $\delta\chi^2$,
which come from distantly mismatched points, is of minor concern for
testing independence.  We are primarily concerned with testing the
null distribution in the region of small $\delta\chi^2 \sim 1$.  The region
of $\delta\chi^2 \sim 1$ is the region of a ``good fit'' between the track
and the source as determined by the relative error.  This is the
region where proposals of correlation might make a difference.  We
chose to examine the integral probability in the bin of $\delta\chi^2 \leq
1$ as our main statistic, similar to the choice of $FB$.  For reference our
figures also show the distribution of $\delta\chi^2$ over the entire range
spanned by the data, and we examine this distribution separately.

We examined the null distribution of $\delta\chi^2$ using two separate
methods to generate uncorrelated tracks and sources.  In one
procedure, we distributed 285 fake source points randomly on the sky,
excluding the galaxy cut, and calculated $\delta\chi^2$ using the
track data set.  In generating the random sample we kept the density
of sources in the northern and southern hemispheres equal to that of
the catalog.  The distribution of $\delta\chi^2$ was determinined by
repeating this 10,000 times.  We call this the ``isotropic null''
(although by excluding the galaxy and using different
northern/southern densities it is not isotropic).  In the other
method we were concerned that correlations or anisotropy of the source
catalogue might skew the isotropic null.  Consequently we took the 285
sources, and generated a new set of 285 sources by 10,000 random
3-parameter orthogonal transformations.  A small fraction of sources
landing inside the galaxy-cut region were re-distributed randomly.
Similarly, any sources landing in the northern or sourthern hemisphere
and causing a density larger than the catalog were randomly
re-distributed into the other hemisphere.  (As a consistency check, we
also generated 10,000 distributions from orthogonal tranformations
without re-distribution, without finding any differences.)
We calculated the $\delta\chi^2$ distribution using the track data
set, and repeated for 10,000 trials.  The ``orthogonal null'' so
determined retains any correlations of the sources among themselves,
while scrambling any relation to the tracks.  Of course both
procedures use the actual track data, so that any bias or correlation
from that is taken into account.  The distribution of $\delta\chi^2$
in the null procedure is shown in Fig.  1. Only one of the null
distribution is shown since the difference between the two nulls was 
found to be small.

A crude summary of the two null distributions finds an expected
average of about 2.5 events, on the average, in the $\delta\chi^2
\leq 1 $ bin.  From the null distribution we can also evaluate the
probability of fluctuations in the independent distribution to give
any number of events above any determined particular value, the
``$P$-value'' or ``confidence level'' of the data.

\section{Results}

The data and actual quasar positions yielded 8 events with $\delta\chi^2 \leq
1 $.  These events can be read off from the Table.  The quasar 3C147
happened to be very close to the boundary of the galaxy cut region and
was excluded from our analysis.  Including 3C147 would exaggerate our
conclusions, and the precise choice of the cut does not change our
results qualitatively.

The null distribution is
convincingly ruled out on the basis of 8 events.  From the null
distribution, the probability for independent source and track
distributions to fluctuate to give 8 events is 0.06\%.  If one were to
assume Gaussian statistics (we do not) the same result would be
expressed as a roughly $3.5\sigma$ effect.  Or, the null hypothesis of
an uncorrelated data sample is ruled out with 99.94 \% confidence
level.

\begin{figure}
\psfig{file=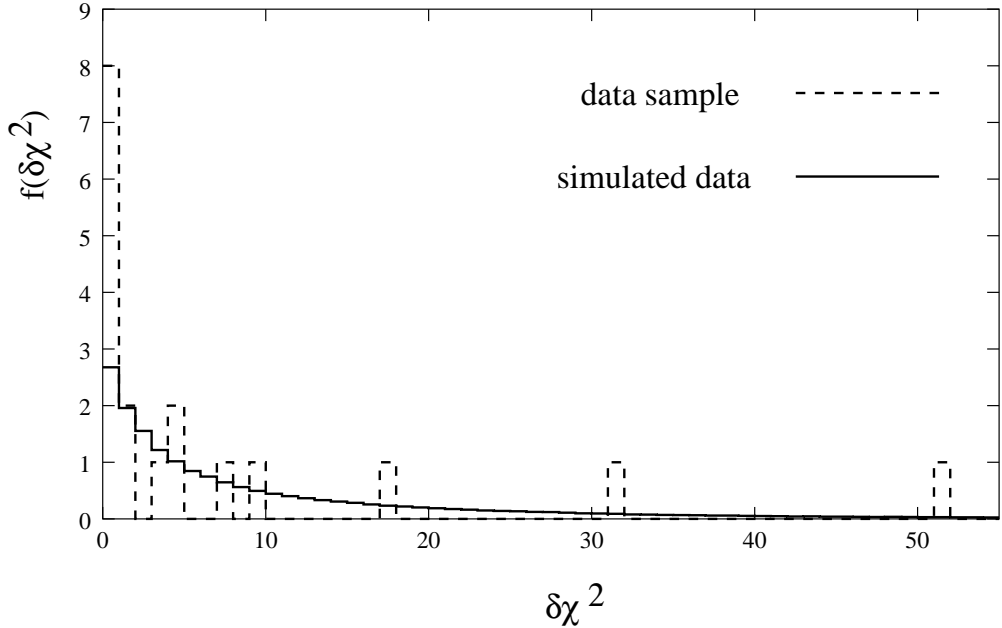} \caption{The distribution of residual $\delta\chi^2$
for individual events for the data set (solid line) and the simulated
data (dashed line).  Both distributions are normalized to the total
number of 18 events cited in the text. }
\label{hist}
\end{figure}

The data's distribution of $\delta\chi^2$ (Fig.  1) shows a peak at
small $\delta\chi^2$ values which is inconsistent with the
distribution of $\delta\chi^2$ seen in the Monte Carlo.  Of the 18
events, 8 have arrival directions within one standard deviation from
one of the compact quasars and 11 have arrival directions within two
standard deviations from one of the compact quasars.  The closest
quasar and its angular separation from the cosmic ray is also shown in
the Table.

As we completed this work, a paper by Sigl {\it et al} appeared
\cite{Sigletal} studying correlations of tracks and selected sources.
Gamma-ray blazars, and a selection of QSO's different from ours leads
to few coincidences in a statistical test described in the references.
There is no contradiction to having a different data set or a
different statistical test lead to no detected correlations, when there are
correlations in our data.  The $P$-values we report are not affected by the
existence of another study involving different assumptions.

For example, we also studied the K\"{u}hr catalogue entries {\it
not} classified as quasars.  This sample of sources with flat or
falling spectrum at low radio frequencies consists of 212 sources.
Among these we found one coincidence with $\delta\chi^2 <1$ and three
coincidences with $\delta\chi^2 <2. $ Applying the same methods as
described above, we found the probability of independence in this case
to be $91\% $.

There appears, then, to be something special about the compact, radio
loud QSO sources.

\begin{figure}
\psfig{file=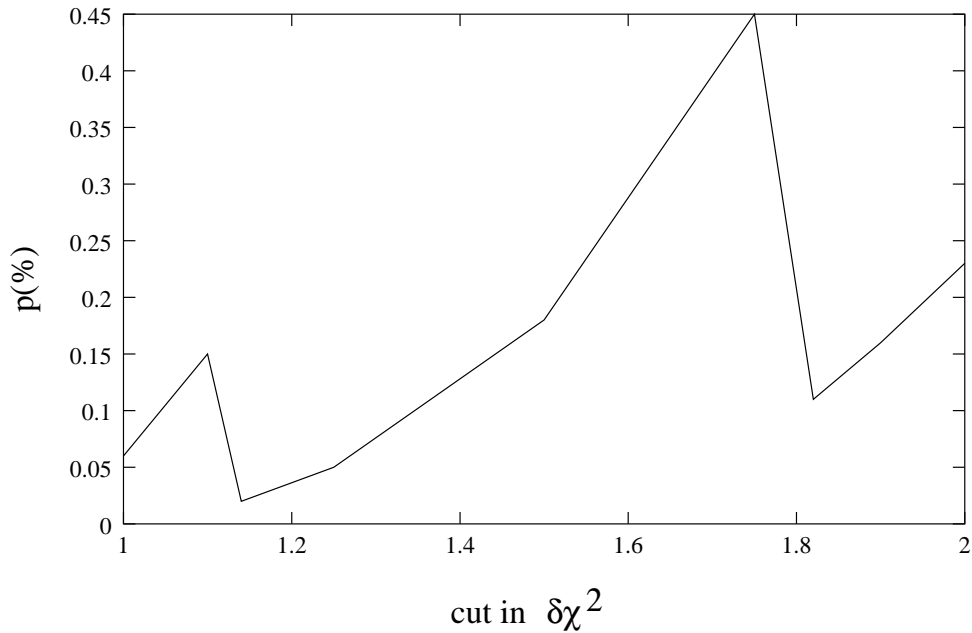} \caption{The probability ($P$-value) $p$\% in
percent that the correlation between the arrival directions of cosmic
rays and compact quasars is due to a chance coincidence,  as a function
of the cutoff on the residual $\delta\chi^2$. }
\label{pvalue}
\end{figure}

There is a third, independent way to estimate the probabilities of the
data observed.  It seems reasonable to consider a  Poisson
distribution for the (integer) number of galaxies inside a given error
cone.  Taking the larger $3^o$ error for
simplicity, each track has been given 0.00382
steradians of solid angle in which to search for a random QSO
coincidence.  There were 285 galaxies spread over $0.74 ( 4 \pi) $
steradians, on the average.  This gives a mean of $ 285 \times
0.00382/( 0.74 ( 4 \pi) )= .12 $ random QSO coincidences per track. 
When we sum
these naively over 18 tracks, and ignore the finite area of the sky,
the estimate gives a mean of about $2.1$ events with $\delta\chi^2 
\sim 1$ in the data
set.  This number is reasonably close to the expectation from the
Monte Carlo of 2.5 events.

In a Poisson distribution, the $P$-value to get 8
events from a distribution with a mean of 2.1 is about $1.5 \times
10^ { -3 }$  This is an independent estimate of the $P$-value of the
data, which we find quite consistent with the Monte Carlo.  Meanwhile the
Monte Carlo results take into account more details.
Indeed, there are
reasons not to trust the Poisson argument: unlike the
case of points dropping on a plane, the sphere has finite area, and is
periodic in its variables, a
constraint that complicates the counting of identical combinations.
When we pursued this in more detail by Monte Carlo, examining the
probabilities of getting $N$ tracks in  regions of fixed angular
size, small deviations from Poisson behavior were observed, which were
difficult to quantify or separate from fluctuations in the absence of
an alternative model.  For the reader who arbitrarily assumes a
Poisson distribution, the confidence level against the null is
about $99.9 \% $.

The procedure so far is sensitive to the precise choice of angular
errors used in the analysis.  We made the choice that was also used in
Ref.  \cite{cluster} for cluster analysis.  To explore this
sensitivity, we examined the dependence of chance probabilities
($P$-values) in the null as a function of the cutoff on the residual
$\delta\chi^2$.  The result is shown in Fig.  \ref{pvalue} where
$P$-values given in percent $p\%$ are shown.  As typical of a small
data set, the plot shows substantial variations in $p$\% as a function
of the cut.  This occurs due to an integer number of track-QSO
coincidences changing as the cut is varied.  Objectively this
constitutes a search with a free parameter favoring the null.  If we
found a case agreeing with the null, we could propose changing the
angular errors as a free parameter, and create support for the null.
The region $0<\delta\chi^2<2$ was searched.  However the maximum $p\%$
found was 0.43\%, so no such arguments are possible.

Finally, in a study to make use of all the $\delta\chi^2$ values in
the data's distribution, we evaluated the formal {\it likelihood} of
data values of $\delta\chi^2$ given the null distribution.  We first
fit the numerically evaluated, normalized orthogonal null distribution
$f_{null}(\delta\chi^2)$ with a smooth interpolating function over the
region $0< \delta\chi^2<60$ units, which is the data's range.  The
smooth fit used several parameters and introduced negligible error.
The log-likelihood of the data ${\cal L}_{data}$ was calculated by
$${\cal L}_{data}=\sum_i log( f_{null}(\chi_i^2).  $$ An advantage of
likelihood is that no binning is involved, and sensitivity to
particular integer counts in each bin is totally eliminated.  We then
calculated the likelihood of a generic competing model, consisting of
a normalized one-sided Gaussian distribution of $ \delta\chi^2 $
centered at zero and with {\it fixed} width $k$ of one unit normalized by
an arbitrary parameter $a$, plus the null distribution normalized by
$1-a$.  It is important that as a model of correlation the generic
model keeps the width of the Gaussian fixed at one unit as part of its
hypothesis.  Since we have no prior information about the relative
populations of correlated and uncorrelated components, parameter $a$
was left free.  Twice the difference of maximum log-likelihoods $2T$
is a very robust statistic distributed by the $\chi_1^2$ distribution,
including the effects of a free-parameter.  The results were $2T=8.8$,
$a=0.41$, which yields a $P$-value rejecting the null at the $99.97\%
$ confidence level.  Dependence on the Gaussian width was also consistent
with $k\sim 1$: the maximum likelihood and 1/2-unit variation occured
at $k=1.24 \pm 0.45$.

\section{Conclusion} For the data set of cosmic rays with energies
above $10^ { 20 } eV$, the hypothesis of statistical independence of
track directions and radio-loud $QSO$'s from the $K\ddot{u}hr$ catlogue is not
consistent with the data.  Several independent statistics yield
results that are consistent with one another and not consistent with
independence.  There exists highly significant correlations between
the track and source directions.

\begin{table}
\begin{tabular}{|c|c|c|c|c|}
\hline
& & &Compact QSO & \\
Cosmic ray event & RA (degrees) & Dec. (degrees)& RA,Dec. (B1950) 
&$\delta\chi^2$ \\
& & & & \\
\hline
\hline
HP101 &  201    &	  	71    & 13 39 29.90, 69 38 30.4 &0.5   \\
HP115&  322     & 	  	47    &   & \\
HP116 &  353      &	 	19    & 23 37 58.3, 26 25 17 & 7.0  \\
HP126 & 179     & 	 	27    & 12 04 54.8, 28 11 34 & 0.8  \\
HP158  &  128     & 	 	67    & 08 59 23.02, 68 09 15.7 & 1.1  \\
HP159    &199     &    	 	44    & 13 25 10.53, 43 42 00.3 & 0.5  \\
YU122 &    74.6     &  	 	45.4  &    &\\
AG101 & 124.25  &		16.8  & 07 59 55.6, 18 18 35& 4.1 \\
AG213 &  18.75   &		21.1  & 01 09 23.7, 22 28 43.0 & 0.97 \\
AG106 &  281.25  &		48.3  & 18 51 08.9, 48 52 06.0 & 0.6   \\
AG144 &  241.5   &		23.0  & 15 38 30.18, 14 57 22.1 & 31.7  \\
AG105 & 298.5   &		18.7  &    &  \\
AG150 & 294.5   &		-5.8  & 20 08 25.90, $-$15 55 37.6 &51.3   \\
AG120 & 349.0  	&		12.3  & 23 28 08.6, 10 43 46  & 4.5  \\
AG104 & 345.75	&		33.9  & 23 27 45.8, 33 31 58 & 9.9  \\
VR135 & 306.7   &		46.8  &   &  \\
SU197 &   187.1   &		32    & 12 19 01.1, 28 30 57 & 1.8   \\
SU155 & 333.1   &		-56   & 22 04 26.2, $-$54 01 14 & 0.6   \\
SU147 & 354.6   &		-74   & 23 53 24.6, $-$68 37 42 & 3.8   \\
SU132 & 117.4   &		-2    & 07 43 20.8, $-$00 37 00 & 0.3   \\
SU126 &  231.3   &		-30   & 15 19 37.6, $-$27 19 29.6& 0.7 \\
SU116 & 356.4   &		-56   & 23 53 24.6, $-$68 37 42 & 17.1	 \\
SU106 & 146.6   &		-43   &  & \\
SU106 & 129.9   &		-26   & & \\
FE320 & $85.2\pm 0.5$		&	$48.0^{+5.2}_{-6.3}$& &	 \\
\hline
\end{tabular}
\caption{Cosmic ray events with energy $E>10^{20}$ eV. For the AGASA events
the error is given by the angular cone radius $\sigma_r=1.8^o$. The
corresponding $\sigma_r$ for the rest of the events excluding the
Fly's Eye event is $3^o$ \cite{cluster}. The blank spaces under the 
compact QSO 
column correspond to cases for which the event was within $\pm 10^o$ of the
galactic plane.}
\end{table}

\bigskip \noindent {\bf Acknowledgements:} We thank P. Biermann for
providing us with the $K\ddot{u}hr$ catalog. This work was supported 
in part under the Department
of Energy grant number DE-FGO2-98ER41079, by the University of Kansas
General Research Fund, and the {\it Kansas Institute for Theoretical
and Computational Science/ K*STAR} program.

\end{document}